\documentclass[aps,pra,twocolumn, notitlepage,10pt,superscriptaddress]{revtex4-1}
\usepackage{graphicx}
\usepackage{times,bbm,amsmath,amssymb}
\usepackage{hyperref}
\usepackage{color}
\usepackage{graphicx}
\usepackage{dcolumn}
\usepackage{bm}
\usepackage{soul}

\usepackage{xcolor}
\usepackage{amssymb}
\usepackage{pifont}
\usepackage{tikz}

\begin{document}

\title{Variational quantum algorithm for experimental photonic multiparameter estimation}

\author{Valeria Cimini}
\affiliation{Dipartimento di Fisica, Sapienza Universit\`{a} di Roma, Piazzale Aldo Moro 5, I-00185 Roma, Italy}

\author{Mauro Valeri}
\affiliation{Dipartimento di Fisica, Sapienza Universit\`{a} di Roma, Piazzale Aldo Moro 5, I-00185 Roma, Italy}

\author{Simone Piacentini}
\affiliation{Istituto di Fotonica e Nanotecnologie, Consiglio Nazionale delle Ricerche (IFN-CNR), Piazza Leonardo da Vinci, 32, I-20133 Milano, Italy}

\author{Francesco Ceccarelli}
\affiliation{Istituto di Fotonica e Nanotecnologie, Consiglio Nazionale delle Ricerche (IFN-CNR), Piazza Leonardo da Vinci, 32, I-20133 Milano, Italy}

\author{Giacomo Corrielli}
\affiliation{Istituto di Fotonica e Nanotecnologie, Consiglio Nazionale delle Ricerche (IFN-CNR), Piazza Leonardo da Vinci, 32, I-20133 Milano, Italy}

\author{Roberto Osellame}
\affiliation{Istituto di Fotonica e Nanotecnologie, Consiglio Nazionale delle Ricerche (IFN-CNR), Piazza Leonardo da Vinci, 32, I-20133 Milano, Italy}

\author{Nicol\`o Spagnolo}
\affiliation{Dipartimento di Fisica, Sapienza Universit\`{a} di Roma, Piazzale Aldo Moro 5, I-00185 Roma, Italy}

\author{Fabio Sciarrino}
\email{fabio.sciarrino@uniroma1.it}
\affiliation{Dipartimento di Fisica, Sapienza Universit\`{a} di Roma, Piazzale Aldo Moro 5, I-00185 Roma, Italy}

\begin{abstract}
Variational quantum metrology represents a powerful tool for optimizing generic estimation strategies, combining the principles of variational optimization with the techniques of quantum metrology. Such optimization procedures result particularly effective for multiparameter estimation problems, where traditional approaches, requiring prior knowledge of the system behavior, often suffer from limitations due to the curse of dimensionality and computational complexity. To overcome these challenges, we develop a variational approach able to efficiently optimize a multiparameter quantum phase sensor operating in a noisy environment. By exploiting the high reconfigurability of an integrated photonic device, we implement a hybrid quantum-classical feedback loop able to enhance the estimation performances, combining classical optimization techniques with quantum circuit evaluations. The latter allows us to compute the system partial derivatives with respect to the variational parameters by applying the parameter-shift rule, and thus reconstruct experimentally the Fisher information matrix. This in turn is adopted as the cost function of a derivative-free classical learning algorithm run to optimize the measurement settings. Our experimental results reveal significant improvements in terms of estimation accuracy and noise robustness, highlighting the potential of the implementation of variational techniques for practical applications in quantum sensing and more generally for quantum information processing with photonic circuits.

\end{abstract}

\maketitle

Variational Quantum Algorithms (VQAs) are emerging as a promising solution to achieve quantum advantage on the currently available Noisy Intermediate-Scale Quantum (NISQ) devices \cite{cerezo2021variational}. These algorithms have been employed for solving a wide range of tasks in different frameworks \cite{zhu2019training} from the estimate of the ground state of a given Hamiltonian \cite{peruzzo2014variational,kandala2017hardware,tilly2022variational}, solving all those problems that undergo the name of variational quantum eigensolver \cite{mcclean2016theory,grimsley2019adaptive,colless2018computation}, for simulating the dynamics of quantum systems \cite{yuan2019theory,endo2020variational}, to quantum error correction problems \cite{xu2021variational,cao2022quantum}. They consist in hybrid classical-quantum algorithms where a classical optimizer is used to minimize a cost function, representative of the solution of the addressed problem, that is efficiently estimated through the quantum system.

Recently, VQAs have been introduced in the context of quantum metrology and sensing considered one of the pillars of current quantum technologies \cite{giovannetti2011advances, polino2020photonic}. The quest for enhanced measurement sensitivities has driven to exploit probe states with quantum correlations in order to go beyond the classical limits. To this end, the optimization of the probe state and the measurement settings are crucial to retrieve such enhanced estimation performances \cite{giovannetti2011advances,paris2004quantum}. Several approaches have been proposed for the optimization of the figures of merit able to certify the quantum estimation performances, such as the Quantum Fisher Information (QFI) \cite{beckey2022variational,yang2022variational}, of probe states in nuclear magnetic resonance systems \cite{liu2022variational,yang2020probe,yang2021hybrid} and of trapped atomic arrays \cite{kaubruegger2019variational,kaubruegger2021quantum} also considering noisy conditions \cite{koczor2020variational} which make the optimization task even harder. However, up to now, the experimental realizations that make use of variational algorithms in the sensing field are still limited to the single-parameter regime.

The next challenge is to apply these methods to the multiparameter regime \cite{szczykulska2016multi,albarelli2020perspective}, where the number of parameters upon which the probe evolution and thus the optimization task depends increases. In such a framework, finding the optimal settings becomes indeed particularly hard and resource expensive. For this reason, several machine learning-based procedures have already been demonstrated as really practical in this scenario \cite{cimini2023deep,cimini2021calibration}. Therefore, the true potential of VQAs can be valued when applied to multiparameter estimation problems allowing to efficiently explore and optimize complex, high-dimensional parameter spaces.
In such a multiparameter framework the simultaneous estimation of the parameters, feasible by exploiting a quantum probe, can be favorable with respect to a separable estimation strategy \cite{PhysRevA.98.012114,PhysRevA.94.052108,PhysRevLett.128.040504}, also in a distributed sensing scenario \cite{liu2021distributed}. However, finding the optimal settings assuring the sensor best performances is harder in this framework, where in general the saturation of the ultimate precision bound, i.e. the quantum Cramér-Rao bound (QCRB), is also not guaranteed \cite{liu2020quantum}. Moreover, when including noise linked to actual experimental conditions, standard approaches can make the required high-dimensional optimization task impossible. Very recently, two theoretical proposals of using VQA in a multiparameter framework have been developed for magnetic field sensing \cite{meyer2021variational,le2023variational}, but their general application on an actual multiparameter sensor is still lacking. 

In this work, we devise and implement a VQA to optimize the operation of a quantum optical multiphase sensor. We extended previous theoretical works \cite{meyer2021variational}, developed using the qubit formalism, to the photonic case devising a methodology tailored for the domain of quantum optics. Such a novel procedure allows us to retrieve the gradient of the experimental quantum photonic circuit depending on the Fock state at the input. The gradient evaluation of the response function of the noisy quantum circuit is crucial to determine its actual estimation performances. This is obtained by applying either the standard parameter-shift rule or the generalized one \cite{mitarai2018quantum, schuld2019evaluating} depending on the probe state used, allowing to retrieve the Fisher information matrix (FIM) from the experimental measurements as a function of the variational parameters, depending on the number of photons in the optical modes. In a consecutive step, a gradient-free learning algorithm updates the parameters in order to optimize the FIM, thus improving the estimation performances. Since the FIM is derived directly from the measured data, all the experimental imperfections will automatically be incorporated into the optimization procedure that thus become inherently resilient to noise.

The device we exploit is an integrated programmable interferometer \cite{wang2020integrated,corrielli2021femtosecond} that can achieve quantum-enhanced performances in the simultaneous estimation of three independent optical phases when injected by two-photon quantum states as demonstrated in \cite{valeri2023experimental}. Thanks to the integrated photonic chip reconfigurability \cite{ceccarelli2020low}, it is possible to tune both the probe preparation and the measurement settings. Notably, the algorithm training and the cost function evaluation are performed directly on the noisy experimental quantum circuit. Therefore, we do not require, at any step of the optimization procedure, the knowledge of the sensor response function, which can remain unknown to its users.

\section*{Results}
\subsection*{Variational Quantum Metrology}

One of the most investigated multiparameter problems is the estimate of a vector of $p$ phases $\vec{\phi}=(\phi_1,...,\phi_p )$ embedded into a multi-arm interferometer \cite{humphreys2013quantum}. The process is investigated by means of a photonic probe prepared in the initial state $\rho_0$ and evolving according to the unitary transformation: 
\begin{equation}\begin{split}
    U_{\vec{\phi}} &= e^{i \sum_{m=1}^p G_m\phi_m}
    \\
    G_m &= n_m
\end{split}\label{eq:generator}
\end{equation}
where $\phi_m$ is the phase shift occurring in the mode $m$ and $G_m$ is the generator of the phase shift transformation, resulting to be the photon number operator of the relative interferometer mode. After the evolution through the studied system, the probe state becomes $\rho_{\vec{\phi}}=U_{\vec{\phi}}\rho_0 U^\dagger_{\vec{\phi}}$. The choice of the probe state plays a crucial role since it determines the ultimate estimation precision achievable.
The second important role is determined in the measurement stage, indeed depending on the implemented positive-operator-valued measure (POVM) $\Pi_x$, it is possible to extract a different amount of information (defined as the classical Fisher information) about the investigated parameters that will be lower or equal to the QFI. 

In the multiparameter scenario, these quantities become matrices and the inequality can be generalized considering their diagonal elements:

\begin{equation}
    \text{Tr}[\text{Cov}(\vec{\phi})] \ge \frac{\text{Tr}[F^{-1}_C(\vec{\phi} )]}{M} \ge \frac{\text{Tr}[F^{-1}_Q(\rho_{\vec{\phi}})]}{M}.
\end{equation}
Here, the first inequality corresponds to the Cramér-Rao bound (CRB), where $\text{Cov}(\vec{\phi})$ is the covariance matrix representing the sensitivity of the estimate, and $M$ represents the number of repetitions of the experiment, while the elements of the FIM are:

\begin{equation}
    F_C(\vec{\phi} )_{ij} = \sum_x \frac{1}{P(x|\vec{\phi})}\Big(\frac{\partial P(x|\vec{\phi} )}{\partial \phi_i}\frac{\partial P(x|\vec{\phi})}{\partial \phi_j} \Big).
\label{eq:FIM}
\end{equation}
Where $P(x|\vec{\phi}) = \text{Tr}\{\Pi_x \rho_{\vec{\phi}}\}$ represents the conditional measurement probability relative to the output $x$. From a practical point of view, to optimize the sensor operation, it is necessary to take into account noise and consider only the possible set of available measurements. Therefore, it is useful to develop a procedure that maximizes the classical Fisher information, finding the optimal settings to reduce the error on the estimate of the vector of parameters $\vec{\phi}$. According to Eq.~\eqref{eq:FIM}, this requires knowing the dynamic of the sensor in order to retrieve the measurement outcome probabilities $P(x|\vec{\phi})$ as well as their derivatives.
Both these tasks are non-trivial since they require to have full knowledge of the sensor operation in a noisy environment. Moreover, it can be hard to find the analytical solution for the optimization of the probe state and the measurement settings considering environmental imperfections in practical applications.

\begin{figure}[ht!]
\centering
\includegraphics[width=0.99\columnwidth]{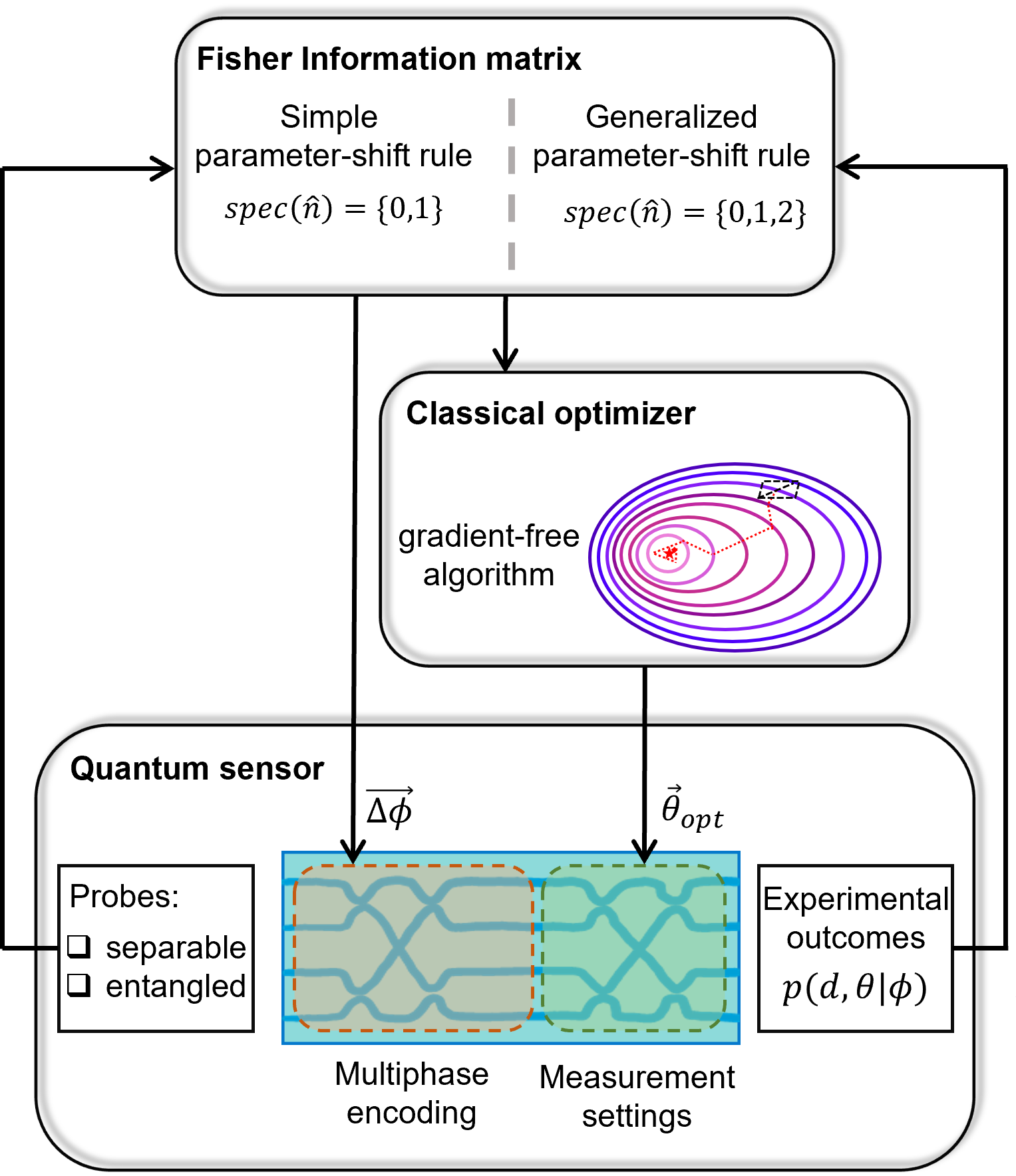}
\caption{\textbf{Scheme of the implemented variational algorithm.} The four-mode integrated interferometer is injected either with single or two-photon probe states. A first layer of internal phase shifters is used to set the triplet of phases $\vec{\phi}$ consisting of the parameters of interest. A second layer of phase shifters can instead be used to perform the optimization by shifting the measurement point of $\vec{\theta}$. Depending on the selected probe state and the set $\vec{\phi}$, the FIM is computed by the quantum circuit through a generalization of the parameter-shift rule. The FIM is then used to compute the cost function of a learning algorithm that optimizes the variational parameter $\vec{\theta}$, enhancing the sensor estimation performances.
}
\label{fig:scheme}
\end{figure}

In particular, we minimize the trace of the inverse of the FIM, chosen as the problem cost function, by optimizing the measurement settings. With this approach, we evaluate the FIM directly with the quantum circuit using the measured data to sample the probability distributions $P(x|\vec{\phi} )$ in Eq.~\eqref{eq:FIM} and, by means of the standard and generalized parameter-shift rules \cite{mitarai2018quantum,schuld2019evaluating}, we compute their partial derivatives with respect to the parameters $\phi_i$.
To perform the minimization procedure we adopted instead a different approach. In order to reduce the number of required experimental points, we implement the function minimization using the Nelder-Mead algorithm \cite{nelder1965simplex}, which results to be one of the most employed and best-performing gradient-free algorithms. 
The scheme of the complete algorithm we have implemented is shown in Fig.\ref{fig:scheme}. 

\subsection*{Gradient of the photonic circuit}

We use the parameter-shift rule \cite{mitarai2018quantum, schuld2019evaluating} to extract the analytic gradient, directly from the experimental data, necessary for the computation of the FIM, as can be seen from Eq.~\eqref{eq:FIM}. Gradient evaluation can be particularly challenging on noisy hardware, thus motivating the application of this rule for quantum machine learning \cite{biamonte2017quantum}. The majority of the optimization procedures indeed require knowledge of the gradient of the cost function. For instance, the most common and powerful optimization algorithm to train machine learning models and neural networks is gradient descent \cite{ruder2016overview}, which uses the gradient of the cost function to determine the model parameter values.
Most of the currently adopted variational quantum algorithms relied either on gradient-free methods or on numerical differentiation, resulting to be highly inefficient or even ineffective. To retrieve the numerical derivative it is indeed necessary to evaluate the function of interest in an infinitesimal shifted point. However, for NISQ devices, the most common scenario corresponds to situations where noise fluctuations are larger than the difference between the function in the original and the shifted point, making it unfeasible to use the finite difference approximation. Moreover, from a practical point of view, numerical differentiation would require having high levels of control of the quantum circuit, allowing one to discriminate among the two settings that differ by an infinitesimal quantity.
The parameter-shift rule solves these issues by obtaining the analytic gradient from the evaluation of the quantum circuit in shifted points of macroscopic size. 

Here, we extend a generalization of the parameter-shift rule (see Methods) to the photonic case by applying the parameter-shift rule to Fock states, retrieving the partial derivatives of the output probabilities with respect to the parameters under study. To our knowledge, no photonic implementation has been realized before obtaining the gradient of the system outcome probabilities allowing its use in a variational framework. In particular, we show that the number of points in which is necessary to evaluate the quantum photonic circuit, in this case, depends on the photon number of the input state.

Considering our quantum system described by Eq.~\eqref{eq:generator}, the generator of the implemented unitary transformation is the photon number operator, therefore, it will depend on the number of photons in the probe state. More specifically, given a probe with $k$ photons, the spectrum of the phase shift generator along the mode $m$ is:
\begin{equation}\label{eq:spec}
    spec(G_m)=\{0,1,2,\dots,k\}.
\end{equation}

We perform the multiparameter estimation of the vector of phases $\vec{\phi}$ using two kinds of probe states: we study the 4-mode interferometer behavior when it is probed with single-photon states and then by exploiting entangled two-photon probes that allow to a achieve superior measurement precision. When the interferometer is injected with single-photon states the Eq.~\eqref{eq:spec} provides $spec(G_m)=\{0,1\}$. Thus, the partial derivatives of the output probability distribution with respect to the three parameters under study, necessary to compute the FIM in Eq.~\eqref{eq:FIM}, can be obtained by applying the simple parameter-shift rule of Eq.~\eqref{eq:par_shift} with $r=\frac{1}{2}$ (see Methods). This allowed us to reconstruct directly from the experimental data the FIM considering the measured outcomes when the device internal phases are set to $\phi_1,\phi_2$ and $\phi_3$ for the reconstruction of the conditional probabilities $P(x|\vec{\phi})$. On the contrary, the measurement performed shifting the phases by a factor $\pm\frac{\pi}{2}$ allows to obtain their derivatives $\partial_{\phi_i} P(x|\vec{\phi})$, also required to reconstruct the FIM [see Eq.~\eqref{eq:FIM}].  More specifically, the partial derivative with respect to the phase $\phi_i$ of the outcome probability is obtained by:

\begin{equation}
    \partial_{\phi_i} P(x|\vec{\phi} ) = \frac{1}{2} \big[ P(x|\vec{\phi} + \vec{\Delta}^{(i)}) -P(x|\vec{\phi} - \vec{\Delta}^{(i)}) \big],
\end{equation}
where $i=1,2,3$, and $\vec{\Delta}^{(i)}$ is a vector with components  $(\vec{\Delta}^{(i)})_j = \frac{\pi}{2}\delta_{ij}$, and $\delta_{ij}$ is the Kronecker's delta.  

The situation changes when we inject into the interferometer two-indistinguishable photons. In this case Eq.~\eqref{eq:spec} gives  $spec\{G_m\}=\{0,1,2\}$, resulting in a generator with more than two distinct eigenvalues. We thus prove that the four-term rule [Eq.~\eqref{eq:gen_par_shift}] can be applied to our system, and we compute the required parameters in order to retrieve the partial derivatives of interest, proving that we obtain the analytic partial derivatives of the measurement outcomes probabilities of the photonic circuit injected with two-photon states:

\begin{equation}
\begin{split}
    \partial_{\phi_i}& P(x|\vec{\phi} )= \big[ P(x|\vec{\phi} + \frac{\pi}{4}\vec{\Delta}^{(i)}) -P(x|\vec{\phi} - \frac{\pi}{4}\vec{\Delta}^{(i)}) \big]+\\
    &-\frac{\sqrt{2}-1}{2} \big[ P(x|\vec{\phi} + \frac{\pi}{2}\vec{\Delta}^{(i)}) -P(x|\vec{\phi} - \frac{\pi}{2}\vec{\Delta}^{(i)}) \big].
\end{split}
\end{equation}

Once having retrieved the analytic gradient of the system's response function, it is possible to compute the FIM and optimize the measurement settings to saturate the CRB. 

\begin{figure*}[htb!]
\centering
\includegraphics[width=0.99\textwidth]{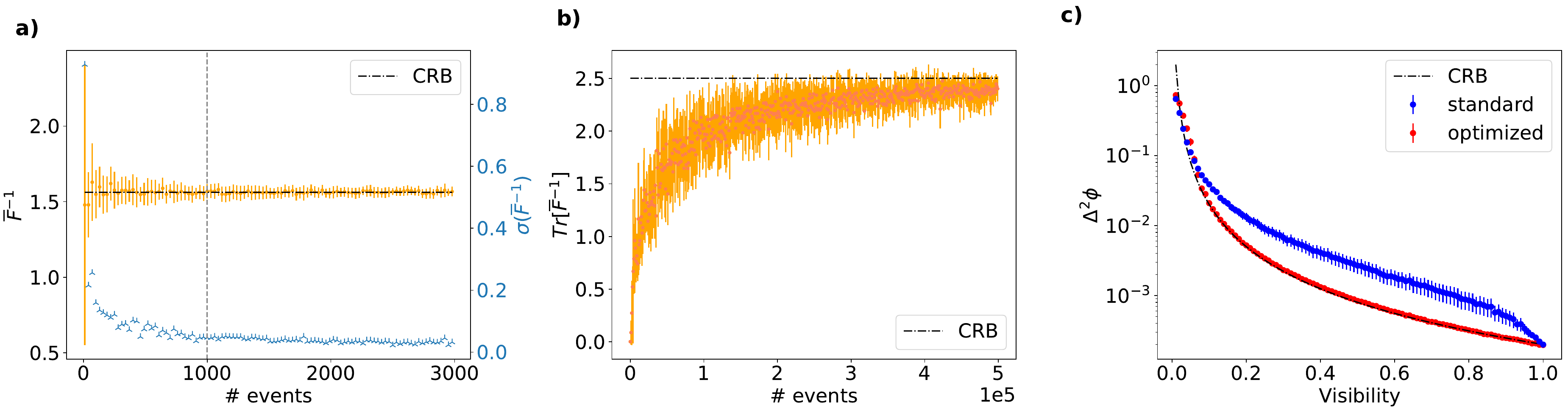}
\caption{\textbf{Numerical simulations addressing the protocol robustness to finite statistics and noise.} Plots in panels \textbf{a)}-\textbf{b)} show how the reliability of the quantum circuit in reconstructing the Fisher information depends on the measurement statistics.
\textbf{a)} Inverse of the Fisher information reconstructed by applying the parameter shift rule to a two-mode interferometer with fringe visibility $v=0.8$. The orange points represent the average results over $30$ different repetitions with a selected random phase $\phi\in [0,\pi]$ with the relative standard deviation also reported as the blu triangles.
\textbf{b)} The same study is done for the four-mode interferometer injected with two-photon states considering the ideal scenario. In this case, the Fisher information is a $3\times 3$ matrix. In both panels, on the x-axis is reported the number of events used to reconstruct the probabilities needed to obtain the Fisher; the dot-dashed lines refer to the respective QCRB.
\textbf{c)} Error on the estimate of $\phi$ in a two-mode interferometer for different levels of noise. Blue points refer to a strategy where the measurement is not shifted while in red the same estimate is done in the shifted point optimized by the learning algorithm which minimize the inverse of the Fisher. All the results are averaged over $15$ phase values each repeated $20$ times. The estimates are performed using a number of probes equal to $5000$. The importance of using the optimized strategy emerges in particular when the level of noise in the system increases.}
\label{fig:statistics}
\end{figure*}

From a practical point of view, it is fundamental to study how the estimate of $P(x|\vec{\phi})$, also referred to as the likelihood function of the system, its partial derivatives, and, as a consequence, the CRB is affected by measurement statistics. We report in Fig.\ref{fig:statistics} such a value applying the parameter-shift rule to Monte Carlo simulated data for systems of increasing dimensionality. More specifically, in Fig.\ref{fig:statistics} a), we report the retrieved CRB of the estimate of a phase shift $\phi$ in a non-perfect (visibility lower than one) two-mode interferometer when the measurement is performed at its most informative point i.e. $\frac{\pi}{2}-\phi$. The robustness of values reconstructed by the parameter-shift rule depends on the number of events $N$ used to sample the probability distributions. In Fig.\ref{fig:statistics} b), we did the same study on the four-arm interferometer when seeded with two-photon states. The increased complexity of the system and of the structure of the probability distributions requires, as expected, a larger number of events to correctly reconstruct the FIM.

The importance of shifting the measurement based on the noise level affecting the probe evolution is demonstrated in Fig.\ref{fig:statistics} c) for a simulated two-mode system. The error on the estimate of the phase shift is compared when measurements are not shifted and when the parameter-shift rule is applied to compute optimal measurement settings.
We perform such a study for different levels of visibility of interference fringes, which is the main source of noise in the system. In the ideal scenario, the Fisher Information does not depend on the value of the parameter under study, therefore there is no gain from the application of the optimization procedure. As soon as a non-ideal value of the visibility is taken into account it becomes evident that, in order to saturate the CRB, it is necessary to optimize the measurement settings even in this simple scenario. Importantly, we demonstrate here that our procedure, outlined in Fig.\ref{fig:scheme}, allows the saturation of the CRB for any value of $V$, where the usefulness of the variational algorithm becomes evident.

\subsection*{Experimental results}

We start by reconstructing experimentally the FIM of our device for each triplet of phases $\vec{\phi}$ investigated. In the experiment, we need to find a balance between the number of events necessary to correctly reconstruct the desired quantities and the overall acquisition time, therefore we fix the number of events for each phase point of this step to $5000$. This number of data is used to reconstruct all the probabilities and their derivatives needed to compute the FIM. Once the FIM is obtained, we minimize the trace of its inverse, which serves as the cost function for the problem.  
This optimization enhances the quality of the estimation, ultimately leading to the saturability of the CRB, in the limit of large resources. Since the figure of merit optimized is reconstructed at each step of the algorithm from the experimental results, it is important to choose an algorithm that converges with a minimal number of function evaluations. 
To mitigate the impact of experimental errors due to finite measurement statistics, we perform the optimization by exploiting a gradient-free algorithm in this second stage. Specifically, the Nelder-Mead algorithm has proven to be particularly effective for our purposes, as it does not rely on the analytic form of the function being minimized. We have also tested other gradient-free optimization algorithms, such as COBYLA, and their performance comparison is provided in the Supplementary Information (SI).

\begin{figure*}[htb!]
\centering
\includegraphics[width=0.99\textwidth]{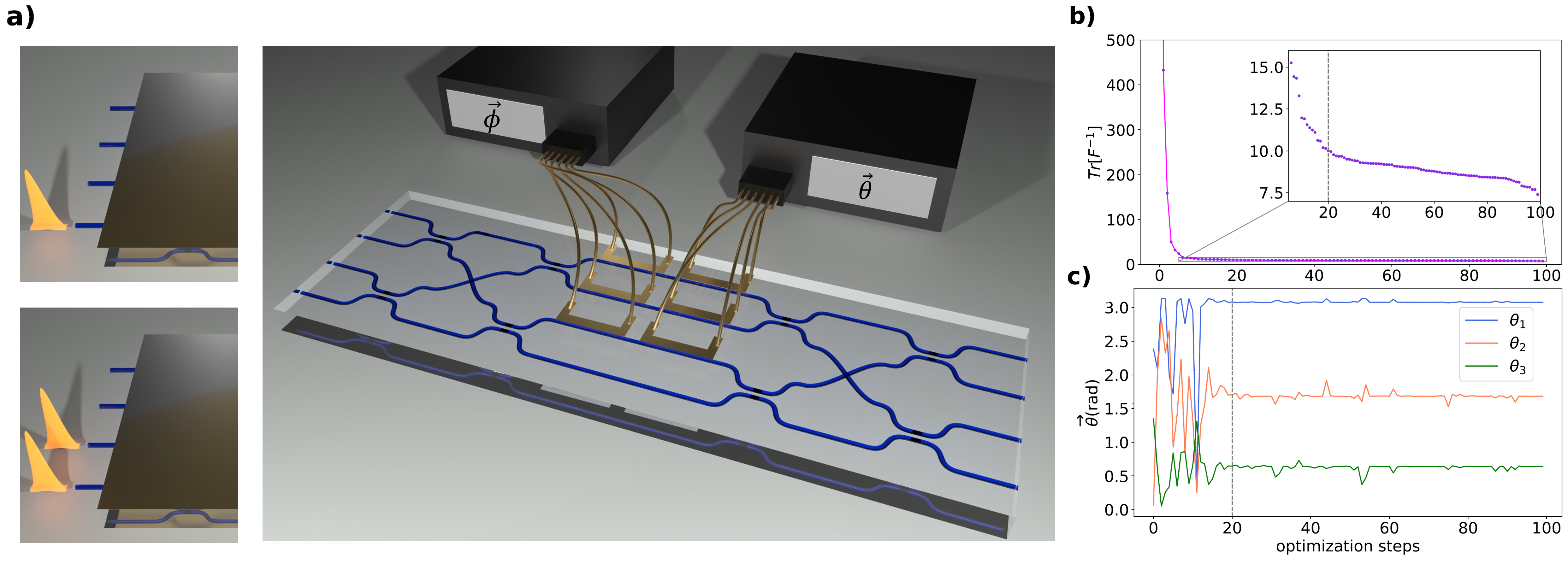}
\caption{\textbf{Integrated photonic circuit and experimental optimization}. \textbf{a)} Sketch of the integrated photonic device and possible selection of different input probe states. The tuned phase shifts $\vec{\phi}$, corresponding to the triplet of phases to estimate, and $\vec{\theta}$, allowing to shift the measurement settings, are highlighted. \textbf{b)} Experimental value of the cost function during the optimization strategy for the configuration $\overline{s}6$ as a function of the optimization steps. \textbf{c)} The Nelder-Mead algorithm minimizes the cost function by changing the variational parameters $\vec{\theta}$, here we report the values set during the optimization procedure for a certain phases triplet.}
\label{fig:optsteps}
\end{figure*}

As described in the Methods and in \cite{valeri2023experimental}, we use the first layer of internal phase shifters to set the three optical phase shifts $\vec{\phi}$, which represents the parameters of interest with respect to the phase of the fourth arm taken as a reference [see Fig.\ref{fig:optsteps} a)]. The second layer of phase shifters is used instead to shift the measurement, thus setting the phases $\vec{\theta}$. Consequently, the overall evolution imparts a phase shift $\vec{\phi}+\vec{\theta}$ to the initial probe. We fix the starting point of the implemented optimization algorithm to $\vec{\theta} =\{\frac{\pi}{2},\frac{\pi}{2},\frac{\pi}{2}\}$ and the number of optimization steps to $m_{fev} = 20$. This choice is justified by looking at Fig.\ref{fig:optsteps} b-c) where the cost function for a certain triplet $\vec{\phi}$ is reported as a function of the algorithm's optimization steps and the related choice of  the vector of parameters $\vec{\theta}$.
Further details can be found in the SI. 

\begin{figure*}[htb!]
\centering
\includegraphics[width=0.99\textwidth]{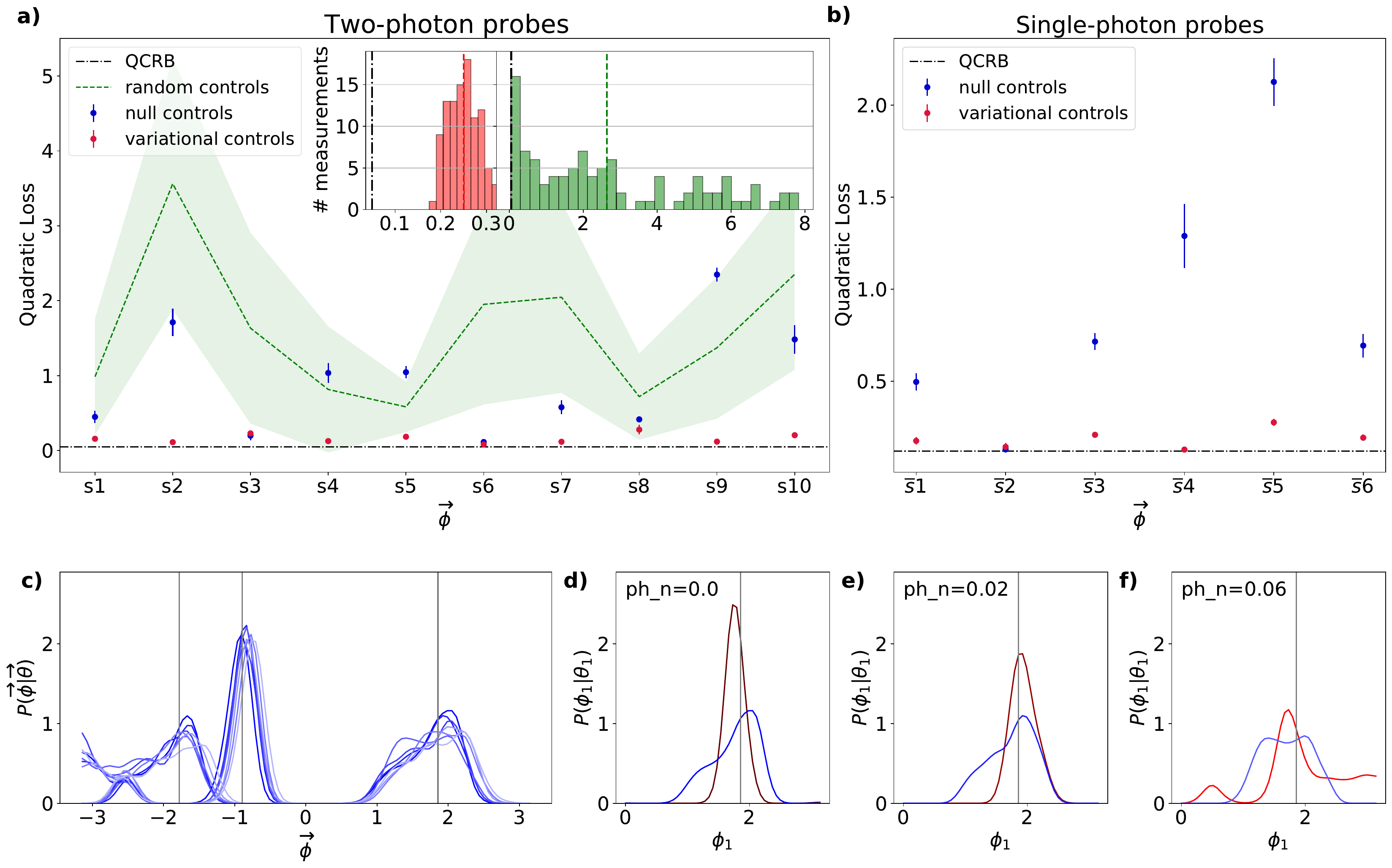}
\caption{\textbf{Experimental estimation performances}.
 Comparison of the estimation performances obtained without adjusting the measurement settings (blu points), shifting them randomly (green dashed line), and setting them to the values retrieved with the developed variational strategy (red points). The results are compared with the QCRB (black dot-dashed line) obtained considering the injected probes in the 4-mode interferometer. The reported experimental results are the average over $30$ different repetitions of the estimate for each phase vector with the relative standard deviation.
\textbf{a)} Experimental quadratic loss obtained using two-photon probes for the estimate of $10$ different phase configurations reported on the x-axis. For the optimization procedure, the FIM is reconstructed by applying the generalized parameter-shift rule.
The inset reports the distributions of the experimental performances achieved on the last triplet i.e. configuration s10 when repeating the estimation procedure $100$ times. 
\textbf{b)} Experimental quadratic loss obtained by injecting single-photon probes in the 4-mode interferometer for the estimate of $6$ different triplets of phases reported on the x-axis. Here, the measurement settings are retrieved by reconstructing the FIM with the standard parameter-shift rule.
\textbf{c)} Averaged experimental reconstructed posterior distributions of the estimate of the configuration $\overline{s}1$ for different phase noise values i.e. $ph_n = (0,0.02,0.04,0.06,0.08,0.1,0.2,0.3)$ achieved with null controls, the dark blue curves represents the results without noise while the lighter the ones with the higher value of noise. \textbf{d)}-\textbf{e)}-\textbf{f)} Comparison of the reconstructed posteriors for the estimate of $\phi_1$ with variational selected controls (red curves) and null controls (blu curves). }
\label{fig:resultmerged}
\end{figure*}

To demonstrate the effectiveness of the optimization algorithm in finding the measurement settings $\vec{\theta}$, we compare the estimation performances obtained through the variational approach with those obtained when $\vec{\theta}$ is set to the null vector or is chosen randomly. We demonstrate the validity of the variational approach independently of the selected phases $\vec{\phi}$ and of the input probe state, by examining the performances achieved with two-photon probes states and with single-photon states.
We quantify the performances of the estimation strategy by looking at the quadratic loss i.e. $(\vec{\phi}-\vec{\phi}_{true})^T(\vec{\phi}-\vec{\phi}_{true})$, where $\vec{\phi}_{true}$ represents the true set values of the phases. We investigate this for a scenario where the interferometer is seeded with $M=50$ probes. For each triplet of investigated phases, randomly chosen in the whole periodicity interval i.e. $\phi_i\in[-\pi,\pi]$, we repeat the procedure $30$ times reporting in Fig.\ref{fig:resultmerged} a-b) the averaged results and the corresponding standard deviations. The $10$ different phase configurations selected for two-photon probes ($s1...s10$) and the $6$ different ones chosen when the system is instead seeded with single-photon probes ($\overline{s}1...\overline{s}6$) are reported in the SI.
The validity of the variational approach in identifying the optimal measurement strategy emerges also from what it is reported in the inset of Fig.\ref{fig:resultmerged} a), where the distribution of results obtained by shifting the measurement to the values computed by the variational algorithm is compared to those achieved by randomly selecting the measurement settings. In the former case, the variability of the results is solely attributed to experimental fluctuations, resulting in a Gaussian distribution. Instead, when the measurement is randomly chosen, the distribution of achieved results reflects the influence of measurement selection on the estimation procedure, leading to a significantly wider spread.

When the interferometer is seeded by two indistinguishable photons, the FIM is computed at every step of the optimization algorithm by applying, as explained above, the generalized parameter-shift rule. 
We apply the same procedure also seeding the interferometer with single-photon states  [Fig.\ref{fig:resultmerged}b)], demonstrating that in this case, the standard parameter-shift rule is instead sufficient to reconstruct the FIM and thus retrieving the optimal measurement settings allowing the saturation of the QCRB.
It is important to note that the QCRB reported in the plots refers to the bound relative to the ideal device corresponding to a QCRB of $2.5/M$ for the two-photon inputs and a QCRB of $6/M$ for single-photon inputs. Therefore, the observed discrepancies are related to the fact that the actual phase sensor has a higher bound due to experimental imperfections, that in general can also depend on the specific triplet of phases under investigation.

Finally, we have experimentally tested the resilience of the developed variational approach to different sources of noise. Specifically, we have examined the performances when introducing additional phase noise $ph_n$ in the single-photon probe estimates and  when reducing the indistinguishability among the two-photon probes.
In Fig.\ref{fig:resultmerged} c) we report the experimental posterior probabilities, reconstructed with the Bayesian estimate \cite{polino2020photonic,valeri2023experimental}, obtained for different noise strengths on the estimate of one triplet of phases without adjusting the measurement settings. The reconstructed posteriors are obtained by averaging the results obtained over 30 different repetitions of the experiment, performed by seeding the system with $M = 500$ single-photon probes. As expected, the performance of the estimation process deteriorates with increasing phase noise, resulting in a dual effect. On one hand, the height of the posterior distribution decreases, and on the other, its mean value results shifted with respect to the true value of the phase. Conversely, if we perform the estimate with the variational approach  under noisy conditions, there is still a significant advantage, even with high levels of noise as shown in Fig.\ref{fig:resultmerged} d-e-f) (see also SI).

\begin{figure*}[htb!]
\centering
\includegraphics[width=0.99\textwidth]{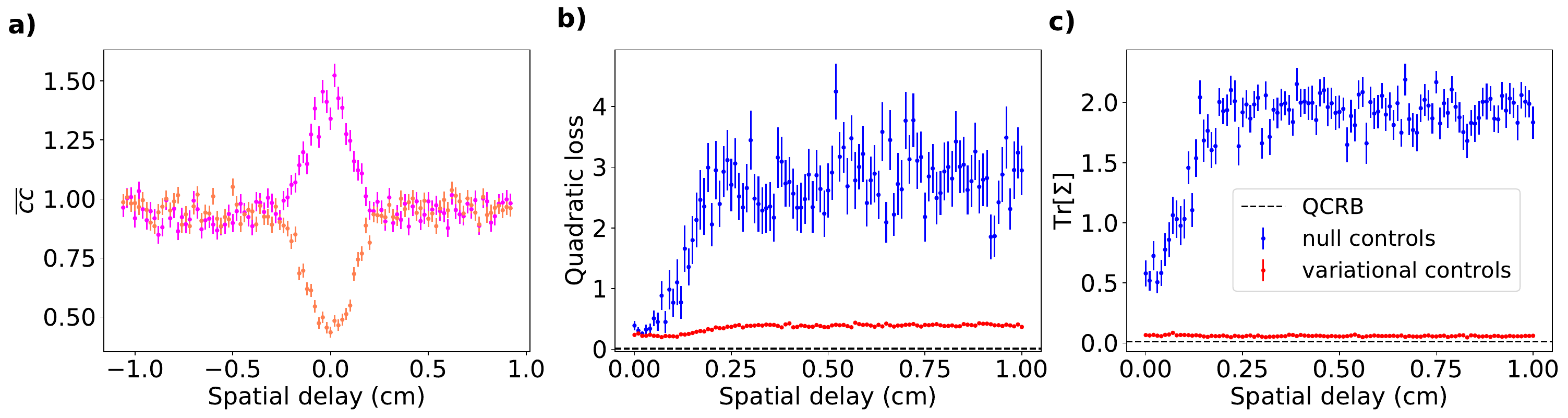}
\caption{\textbf{Experimental results as a function of degree of two-photon indistinguishability.} \textbf{a)} Normalized coincidences events among different output modes registered when changing the respective arrival time of the two photons. This is changed by moving a translation stage in one of the photon paths, allowing to adjust the degree of indistinguishability of the two photons. The magenta points refer to events with both photons in the first output while orange points to events with one photon on the first and one on the second output. \textbf{b)}-\textbf{c)} Study of the estimation performances in terms of quadratic loss and variances, respectively, obtained when degrading the indistinguishability of the two photons. The blu points are the results obtained without shifting the measurement while the red points are the ones achieved with the variational strategy. The results are the avereges of the estimate with $M=200$ two-photon probes of one triplet of phases repeated $30$ times.}
\label{fig:hom}
\end{figure*}

Another investigated scenario consists in studying the effectiveness of the variational algorithm when reducing the degree of indistinguishability of the two-photon probes. In this situation, the estimation precision deteriorates losing the quantum-enhanced performances achieved using entangled probes. As a consequence, the ultimate precision bound increases. For the ideal device, the QCRB obtained when injecting into the system $M$ indistinguishable photon pairs is $2.5/M$, while it becomes $3/M$ for completely distinguishable photons. It follows that the estimation performances will get worse when increasing the level of distinguishability. In these conditions, it is interesting to study how the estimation performances decay and compare the results that can be achieved when the measurements are optimized through the variational approach in this noisy scenario. In Fig.\ref{fig:hom} we show how the level of indistinguishability can be tuned through a delay line, monitoring the bunching and anti-bunching effects at the outputs of the chip [Fig.\ref{fig:hom} a)], while in Fig.\ref{fig:hom} b-c) we report the achieved performances in terms of quadratic loss and variances, respectively, for the estimate of a triplet of phases. Notably, even though a slight reduction of the performances is observed, the enhancement achieved with the variational strategy, with respect to the non-shifted measurements scenario, becomes even more pronounced when the two photons become completely distinguishable.

\section*{Conclusions}

In this work, we have implemented a variational approach to optimize a multiparameter quantum phase sensor operating in a noisy environment. Rather than relying on classical computations to determine the optimal measurement settings, a procedure that can be particularly hard, especially in the multiparameter framework, our method exploits directly the quantum circuit to reconstruct a meaningful cost function successively fed into a classical optimization algorithm. 

We demonstrated the validity of our method by experimentally reconstructing the FIM and optimizing the measurement settings for a multiphase sensor probed with single-photon and entangled two-photon probe states. 
This approach allows us to overcome the limitations of traditional methods and our experimental results showed significant improvements in estimation accuracy and noise robustness compared to cases where the measurement settings were not optimized or chosen randomly. The variational approach is resilient to noise and can effectively explore and optimize the high-dimensional multi-parameter spaces, making it a promising tool for practical applications in quantum sensing and quantum information processing with photonic circuits.

Here, we extended and validate experimentally the procedure presented in \cite{meyer2021variational} to the photonic case, developing a method that, depending on the selected probe state, and thus on the number of photons in the optical modes, retrieves 
the circuit gradient directly from the measurement outcomes. Such a method can benefit in general quantum photonic protocols requiring gradient evaluation.

In conclusion, the obtained results pave the way for the implementation of variational techniques in the challenging multiparameter framework and set the stage for enhancing both quantum sensing capabilities and for future advancements in quantum information processing with photonic circuits. 

\section*{Methods}

\subsection*{Standard and generalized parameter-shift rule} 
The parameter-shift rule, as demonstrated in \cite{schuld2019evaluating,mitarai2018quantum,PhysRevLett.118.150503,banchi2021measuring}, provides a method to obtain the partial derivatives of quantum expectation values with respect to a circuit parameter $x$ directly by using the quantum hardware as follows: 
\begin{equation}
   \partial_x \langle A\rangle = r [\langle A\rangle\big(x+\frac{\pi}{4r}\big)-\langle A\rangle\big(x-\frac{\pi}{4r}\big)],
\end{equation}
where $\langle A\rangle = \langle 0| U^\dagger(x) A U(x) |0 \rangle, $  with $U(x)$ representing the evolution obtained applying the overall set of gates that make up the quantum circuit, and $r=\frac{1}{2}$. The validity of such a rule has been first demonstrated for gates with generators with two unique eigenvalues such as single-qubit rotations, linear combinations of Pauli operators, and for Gaussian gates in the continuous variable regime \cite{schuld2019evaluating}.
This has also been extended to unitary evolution of the form $U_\rho (\phi) = e^{i \phi G} \rho e^{-i \phi G}$ depending on a single parameter $\phi$ and on the generator $G$, whose spectrum has at most two distinct eigenvalues: $spec(G)=\{\lambda_1,\lambda_2\}$, obtaining:
\begin{equation}
    \partial_\phi U_\rho(\phi) = r\big[U_\rho(\phi+\frac{\pi}{4r})-U_\rho(\phi-\frac{\pi}{4r})\big],
\label{eq:par_shift}    
\end{equation}
with $r=\frac{|\lambda_1-\lambda_2|}{2}$. 
Although the rule has initially been derived to retrieve the gradient of quantum unitary evolutions, it has been recently extended to evolutions affected by noise, proving its validity also after the application of dephasing and depolarizing channels \cite{PhysRevLett.118.150503,meyer2021variational} and for multi-qubit quantum evolution using stochastic methods \cite{banchi2021measuring}. 
Then, this method has been generalized to generators with a larger spectrum. In this case, it is possible to extend the rule either by exploiting a polynomial expansion of the unitary transformation \cite{izmaylov2021analytic}, or using trigonometric functions\cite{wierichs2022general}, or through spectral decomposition of the generator \cite{kyriienko2021generalized}.
In particular, for a generator whose spectrum has three distinct but equidistant eigenvalues in \cite{anselmetti2021local} has been demonstrated that the gradient evaluation requires four evaluations of $U_\rho(\phi)$ at different points:
\begin{equation}
\begin{split}
    \partial_\phi U_\rho(\phi) &= r_1\big[U_\rho(\phi+x_1)-U_\rho(\phi-x_1)\big]+\\
    &-r_2\big[U_\rho(\phi+x_2)-U_\rho(\phi-x_2)\big].
    \end{split}
\label{eq:gen_par_shift}    
\end{equation}
(see SI for the details). 
\subsection*{Photon source}
The single- and two-photon probe states employed are generated by a non-collinear spontaneous parametric down-conversion source of Type I. In particular, photon pairs at 808nm are emitted by the source, which are then coupled into single-mode fibers. For the study with single-photon states one photon is directly detected by a single-photon avalanche photodiode, acting as a trigger, while the other is injected into the integrated circuit. For the two-photon probes scenario both the photons are injected into the chip after being made indistinguishable in polarization and time of arrival degrees of freedom through wave plates and a delay line. To properly address all the possible outcome configurations, in this scenario, we use 4 fiber beam splitters for each of the outcomes of the circuit in turn connected to 8 single-photon avalanche photodiode detectors.
\subsection*{Integrated photonic device}
The integrated circuit consists of two sequentially arranged sections, each made up of four directional couplers set up in a dual-layer arrangement and a three-dimensional waveguide intersection. The different optical phases are obtained by applying voltages on microheaters that allow setting specific phase shifts in the desired optical modes.
The interferometric area between the two sections consists of four straight waveguide segments, the optical phases of which $\vec{\phi}+\vec{\theta}$ can be adjusted using eight thermal phase shifters. The total length of the device is 3.6 cm. All thermal shifters were created using femtosecond laser micromachining and include laser-etched isolation trenches around each microheater \cite{ceccarelli2020low}. Lastly, two four-channel single-mode fiber arrays have been glued at the input and output facets of the interferometer, with average total insertion losses (from the connector of the input fiber to the connectors of the output fiber array) of 2.5 dB (insertion loss of the bare device before pigtailing of 1.5 dB).

\section*{Acknowledgments}
This work is supported by the ERC Advanced Grant QU-BOSS (QUantum advantage via non-linear BOSon Sampling, Grant Agreement No. 884676) and by the PNRR MUR project PE0000023-NQSTI (Spoke 4 and Spoke 7). N. S. would like to acknowledge funding from Sapienza Università di Roma via Bando Seed PNR 2021, Project AQUSENSING (Advanced Calibration and Control of Quantum Sensors via Machine Learning).
The integrated circuit was partially fabricated at PoliFAB, the micro- and nanofabrication facility of Politecnico di Milano. F. C. and R. O. would like to thank the PoliFAB staff for the valuable technical support.

\providecommand{\noopsort}[1]{}\providecommand{\singleletter}[1]{#1}%

\end{document}


\title{Supplementary information for: Variational quantum algorithm for experimental photonic multiparameter estimation}

\author{Valeria Cimini}
\affiliation{Dipartimento di Fisica, Sapienza Universit\`{a} di Roma, Piazzale Aldo Moro 5, I-00185 Roma, Italy}

\author{Mauro Valeri}
\affiliation{Dipartimento di Fisica, Sapienza Universit\`{a} di Roma, Piazzale Aldo Moro 5, I-00185 Roma, Italy}

\author{Simone Piacentini}
\affiliation{Istituto di Fotonica e Nanotecnologie, Consiglio Nazionale delle Ricerche (IFN-CNR), Piazza Leonardo da Vinci, 32, I-20133 Milano, Italy}

\author{Francesco Ceccarelli}
\affiliation{Istituto di Fotonica e Nanotecnologie, Consiglio Nazionale delle Ricerche (IFN-CNR), Piazza Leonardo da Vinci, 32, I-20133 Milano, Italy}

\author{Giacomo Corrielli}
\affiliation{Istituto di Fotonica e Nanotecnologie, Consiglio Nazionale delle Ricerche (IFN-CNR), Piazza Leonardo da Vinci, 32, I-20133 Milano, Italy}

\author{Roberto Osellame}
\affiliation{Istituto di Fotonica e Nanotecnologie, Consiglio Nazionale delle Ricerche (IFN-CNR), Piazza Leonardo da Vinci, 32, I-20133 Milano, Italy}

\author{Nicol\`o Spagnolo}
\affiliation{Dipartimento di Fisica, Sapienza Universit\`{a} di Roma, Piazzale Aldo Moro 5, I-00185 Roma, Italy}

\author{Fabio Sciarrino}
\email{fabio.sciarrino@uniroma1.it}
\affiliation{Dipartimento di Fisica, Sapienza Universit\`{a} di Roma, Piazzale Aldo Moro 5, I-00185 Roma, Italy}

\maketitle

\section{Derivation of the generalized parameter shift rule}

In order to retrieve the parameters needed to apply the parameter-shift rule to our system we start considering an evolution $U(\varphi) = e^{i\phi G}$. The function of interest is the expectation value on the evolved state:
\begin{equation}
    f(\varphi) = \langle U(\varphi)^\dagger A U(\varphi)\rangle,
\end{equation}
and its derivative will be given by:
\begin{equation}
   \frac{\partial}{\partial\varphi}f(\varphi) = \langle U(\varphi)^\dagger \big(i [A,G]\big) U(\varphi)\rangle.
\end{equation}
Following the results presented in \cite{anselmetti2021local} we expand our unitary evolution using the exponential series and considering a generic gate with spectrum $\{-1,0,1\}$ obtaining:
\begin{equation}
    U(\alpha)^\dagger A U(\alpha)-U(-\alpha)^\dagger A U(-\alpha) = 2i \sin\alpha[A,G]+i(\sin2\alpha-2 \sin\alpha)[G,AGA].
\end{equation}
This relation can be used to retrieve the commutator, needed to compute the derivative, by studying the system in a second angle $\beta$. At this point, the commutator can be obtained using a linear combination of the results obtained in the four points $\pm\alpha$ and $\pm\beta$:
\begin{equation}
    i [A,G] = \big[d_1 (2i\sin\alpha[A,G])- d_2 (2i\sin\beta[A,G])\big]
     \big[d_1(i\sin 2\alpha -2i\sin\alpha)-d_2(i\sin 2\beta -2i\sin\beta)\big] =0.
\end{equation}
Therefore, the necessary coefficients can be computed by solving the following equations;
\begin{equation}
\begin{split}
    &d_1\sin\alpha-d_2\sin\beta = \frac{1}{2}\\
    & d_1\sin2\alpha-d_2\sin2\beta = 1.
    \end{split}
\end{equation}
It is important to stress that this still holds in our case even if the spectrum of the generator is $\{0,1,2\}$. Indeed, in \cite{anselmetti2021local} the authors demonstrate that the four-term parameter-shift rule works for any gate with spectrum $\{-a+c,c,a+c\}$ where $c$ does not influence the gradient since it is absorbed as a global phase, while $a$ is absorbed in the variational parameter $\varphi$.

Given our evolution, we then compute the required parameters
that result to be:

\begin{equation}
    d_1 = 1, \qquad \alpha =\frac{\pi}{4}, \qquad d_2=\frac{\sqrt{2}-1}{2}, \qquad \beta=\frac{\pi}{2}.
\end{equation}

\section{Nelder-Mead optimization algorithm}

The Nelder-Mead algorithm \cite{singer2009nelder} is an iterative optimization gradient-free method used for the minimization of an objective function in a multidimensional space. Indeed, it does not require the computation of derivatives of the objective function making it suitable for optimizing functions that are not easily differentiable or when the derivative information is not available.

The algorithm begins by constructing an initial simplex, formed by a set of vertices representing points in the search space. The simplex is initially chosen based on the initial guess or exploration of the problem domain.

In each iteration, the algorithm evaluates the objective function at the vertices of the simplex and determines the best (lowest) and worst (highest) function values. These evaluations provide information about the landscape of the objective function and guide the subsequent steps. The simplex is updated consequently through a series of transformations: reflection, expansion, and contraction.
The first involves reflecting the worst vertex through the centroid of the remaining vertices. If the reflected point has a better function value than the second-worst vertex but is worse than the best vertex, the algorithm accepts this reflection and replaces the worst vertex with the reflected point. This step allows the algorithm to move towards a better region of the search space.
If the reflected point has the best function value observed so far, the algorithm may perform an expansion. The expansion involves stretching the simplex further in the direction of the reflection to explore the search space more extensively. If the expanded point has a better function value than the reflected point, it replaces the worst vertex; otherwise, the reflected point is kept.
On the contrary, contraction involves shrinking the simplex towards the best vertex. The contracted point is evaluated, and if it has a better function value than the worst vertex, it replaces the worst vertex.

The Nelder-Mead algorithm continues iterating, updating and refining the simplex until a termination condition is met, such as reaching a maximum number of function evaluations, achieving a desired level of precision, or satisfying certain convergence criteria. The final vertex of the simplex represents the solution that minimizes the objective function.
Therefore, it is important to note that the Nelder-Mead algorithm is sensitive to the choice of initial simplex and other parameters. 
In our algorithm, we initialize randomly the vertices of the initial simplex of the Nelder-Mead algorithm that, given the dimensionality of our problem, corresponds to a tetrahedron. Given the initial randomness, the algorithm after $ fev$ evaluations can end up in different minima, therefore, in order to increase the probability to find a global minimum, we repeat the optimization strategy $3$ times taking as the final vector of parameters $\vec{\theta}_{opt}$ the one associated to the lowest value of the chosen cost function.

Here, through some numerical simulations, we justify the choices done on the initial simplex and the number of maximum function evaluations which from an experimental point of view require to be the minimum value granting a high probability to arrive at the global minimum of the objective function.

\begin{figure}[ht!]
\centering
\includegraphics[width=0.9\textwidth]{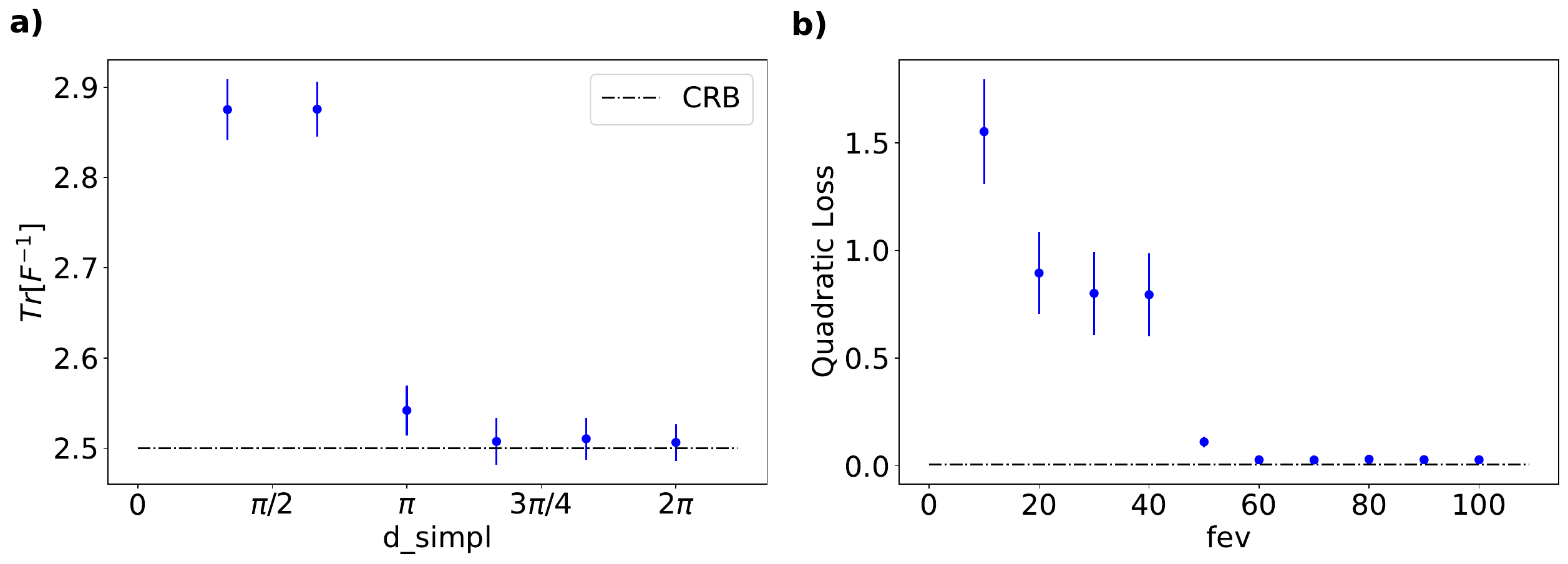}
\caption{\textbf{a)} Values of the CRB obtained performing the measurement in the point retrieved with the Nelder-Mead algorithm run with different sizes of the initial simplex. The results correspond to the averages and the relative standard deviations obtained reconstructing the FIM for $100$ different random phase triplets fixing the maximum number of function evaluations to $100$. \textbf{b)} Quadratic loss among the estimate triplet and the true values obtained when performing the measurement in the shifted point retrieved with the Nelder-Mead minimization run with the maximum number of function evaluations reported on the x-axis. The performances are the average over $100$ different phase triplets each repeated $10$ times obtained with $50$ two-photon probes.
}
\label{fig:maxfev}
\end{figure}

We start by testing its performances on simulations with the ideal device outcome probabilities. Applying the parameter-shift rule, we reconstruct the Fisher Information matrix (FIM) and its inverse from these probabilities. The trace of its inverse is then used as the cost function of the Nelder-Mead algorithm that optimizes the measurement strategy in order to minimize the cost function. We study in Fig.\ref{fig:maxfev} the results of the Nelder-Mead algorithm in minimizing the trace of the inverse of the FIM as a function of the dimensionality of the initial simplex [$d_{\text{simpl}}$ in Fig.\ref{fig:maxfev} a)] and the estimation performances achieved using a different number of function evaluations (fev) in the minimization procedure [Fig.\ref{fig:maxfev} b)].

\begin{figure}[ht!]
\centering
\includegraphics[width=0.9\textwidth]{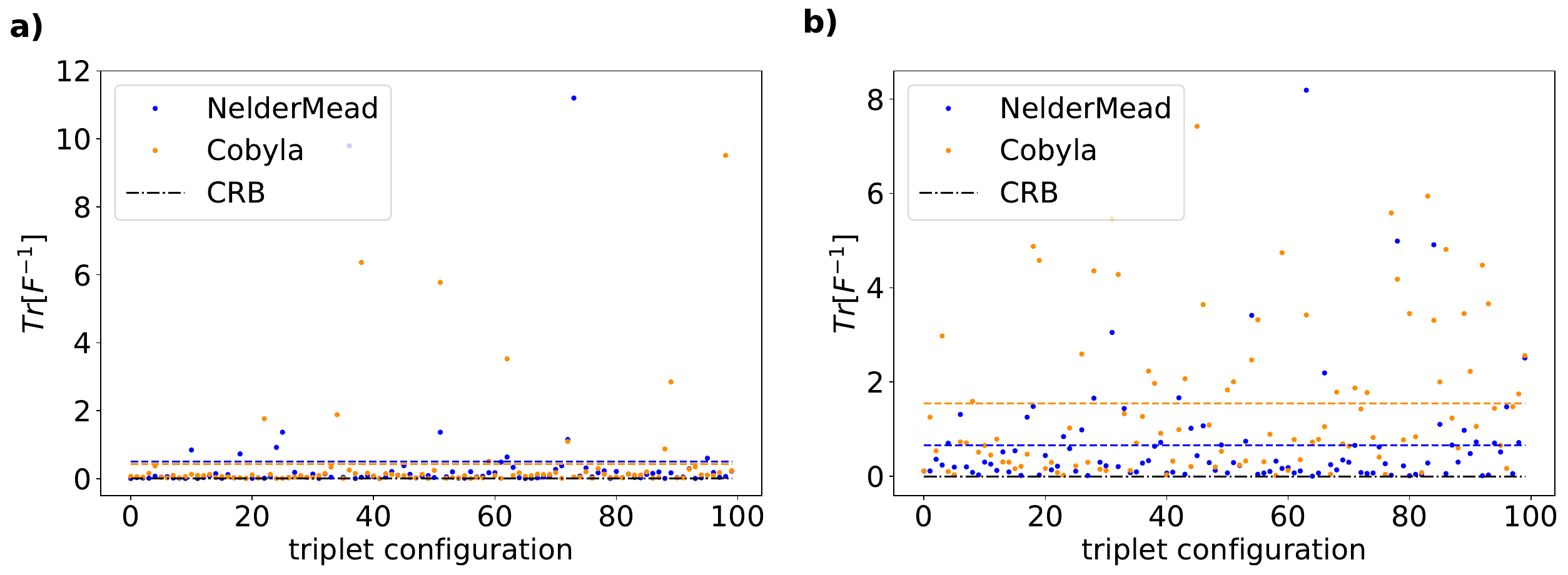}
\caption{\textbf{a)} Comparison of the results achieved when minimizing the trace of the inverse of FIM, reconstructed applying the parameter-shift rule to the functional form of the outcome probabilities of the ideal device injected with two-photon probes, with the Nelder-Mead algorithm (blue points) and with COBYLA (orange points) for $100$ different random phase triplets configurations ($\vec{\phi}$). The dashed lines are the averages of these results with the same color code. \textbf{b)} The same comparison is done when the FIM is reconstructed without knowing the probabilities functional form but directly on Monte Carlo simulated data.
}
\label{fig:cob}
\end{figure}

In order to properly select the suitable optimization algorithm we have also performed a comparison of the performances achieved with the Nelder-Mead minimization with the ones obtained with another of the most employed optimization algorithms, namely COBYLA \cite{powell2007view}. 
It is another derivative-free optimization technique, approximating the function to minimize using linear models and iteratively updating the search direction to improve the approximation and converge toward the optimal solution. One of its strengths is its efficiency compared to Nelder-Mead, indeed it usually requires fewer function evaluations to complete the minimization task.
The results achieved with these two algorithms when reconstructing the FIM applying the parameter-shift rule using the functional form of the ideal probabilities are comparable [Fig.\ref{fig:cob} a)]. However, when we do not use the functional form of the probabilities but instead we reconstruct the FIM using Monte Carlo simulated data the two optimizations give different results. In this case, that is the one relevant from the experimental point of view, the Nelder-Mead minimization results are on average closer to the CRB making this optimization more robust to statistical noise.

\section{Details on experimental measurements}
\label{sec:details}

The different triplet configurations investigated, representing different combinations of phase shifts,  have been selected randomly by applying different voltages to the three phase shifters of the first-layer on the internal arms of the integrated device. 
In the following tables, we report the values of the phases $\vec{\phi}$ relative to the studied configurations for both the two-photon probes scenario and the single-photon one:

\begin{table}[htb]
    
    \begin{subtable}[t]{.5\textwidth}
        

\caption{Triplet configurations investigated with two-photon probes}

\begin{tabular}{|c|c|}
\hline
\multicolumn{2}{|c|}{\textbf{Two-photon probes}} \\
\hline
\hline
Configuration & $\vec{\phi}$  \\
\hline
$s1$ & $[-0.588,  1.302,  0.511]$ \\
$s2$ & $[-0.636,  1.379, -0.024]$ \\
$s3$ & $[-0.153,  0.902,  0.587]$\\
$s4$ & $[0.830, 2.263 , 0.703]$ \\
$s5$ & $[-0.723,  1.938, -1.241]$ \\
$s6$ & $[0.863 , 1.132, 0.111]$ \\
$s7$ & $[-0.617,  1.037, -0.369]$ \\
$s8$ & $[1.498, 0.776, 0.571]$ \\
$s9$ & $[0.777, 2.330, 0.204]$ \\
$s10$ & $[ 0.210,  1.436, -0.150]$ \\
\hline
\end{tabular}
    \end{subtable}%
   \begin{subtable}[t]{.5\textwidth}
\caption{Triplet configurations investigated with single-photon probes}

        \begin{tabular}{|c|c|}
\hline
\multicolumn{2}{|c|}{\textbf{Single-photon probes}} \\
\hline
\hline
Configuration & $\vec{\phi}$  \\
\hline
$\overline{s}1$ & $[ 1.856, -1.782, -0.896]$ \\
$\overline{s}2$ & $[ 0.748,  2.741, -2.225]$ \\
$\overline{s}3$ & $[0.768,  2.636, -1.582]$\\
$\overline{s}4$ & $[0.701,  2.452, -0.978]$ \\
$\overline{s}5$ & $[0.796, 2.322, 0.160]$ \\
$\overline{s}6$ & $[0.834, 2.277, 0.706]$ \\

\hline
\end{tabular}
    \end{subtable}
\end{table}

With the aim of reducing the impact of experimental fluctuations on the developed methodology, we focus here on one configuration of the two scenarios. For those configurations we repeat the measurements necessary to reconstruct the FIM, $30$ times in order to study the variation of the experimental CRB for different measurements, providing a comprehensive understanding of the estimation performance.
For single-photon probes, such a study has been done both without shifting the measurement settings i.e. $\vec{\theta}=(0,0,0)$ and when setting them to the optimal point retrieved with the variational approach. The obtained experimental results are reported in Fig.\ref{fig:FIrep}.

\begin{figure}[ht!]
\centering
\includegraphics[width=0.9\textwidth]{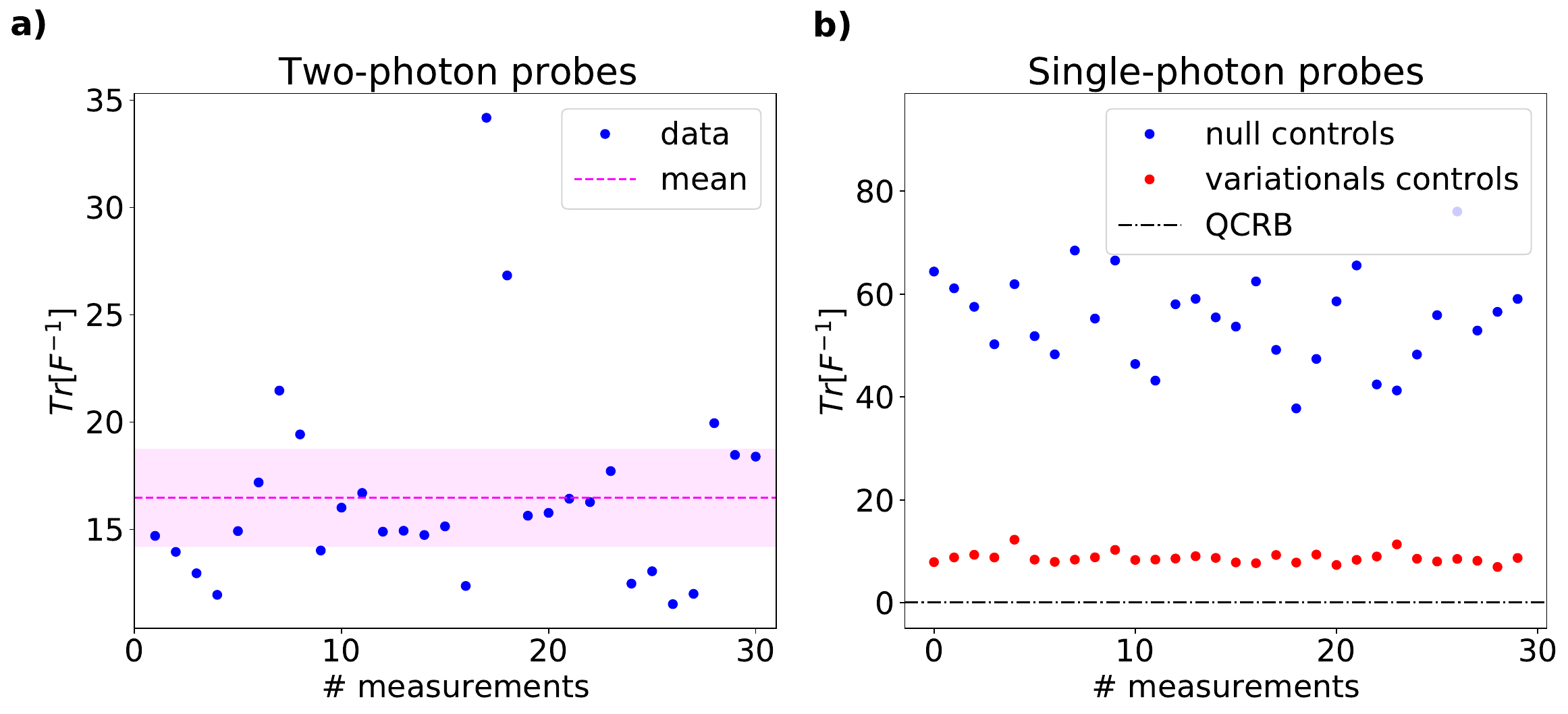}
\caption{Trace of the inverse of the experimental reconstructed FI measured multiple times. \textbf{a)} Results obtained reconstructing each probability with $5000$ events for the configuration s5. \textbf{b)} Results obtained reconstructing each probability with $5000$ events for the configuration $\overline{s}1$.
}
\label{fig:FIrep}
\end{figure}

\begin{figure}[ht!]
\centering
\includegraphics[width=\textwidth]{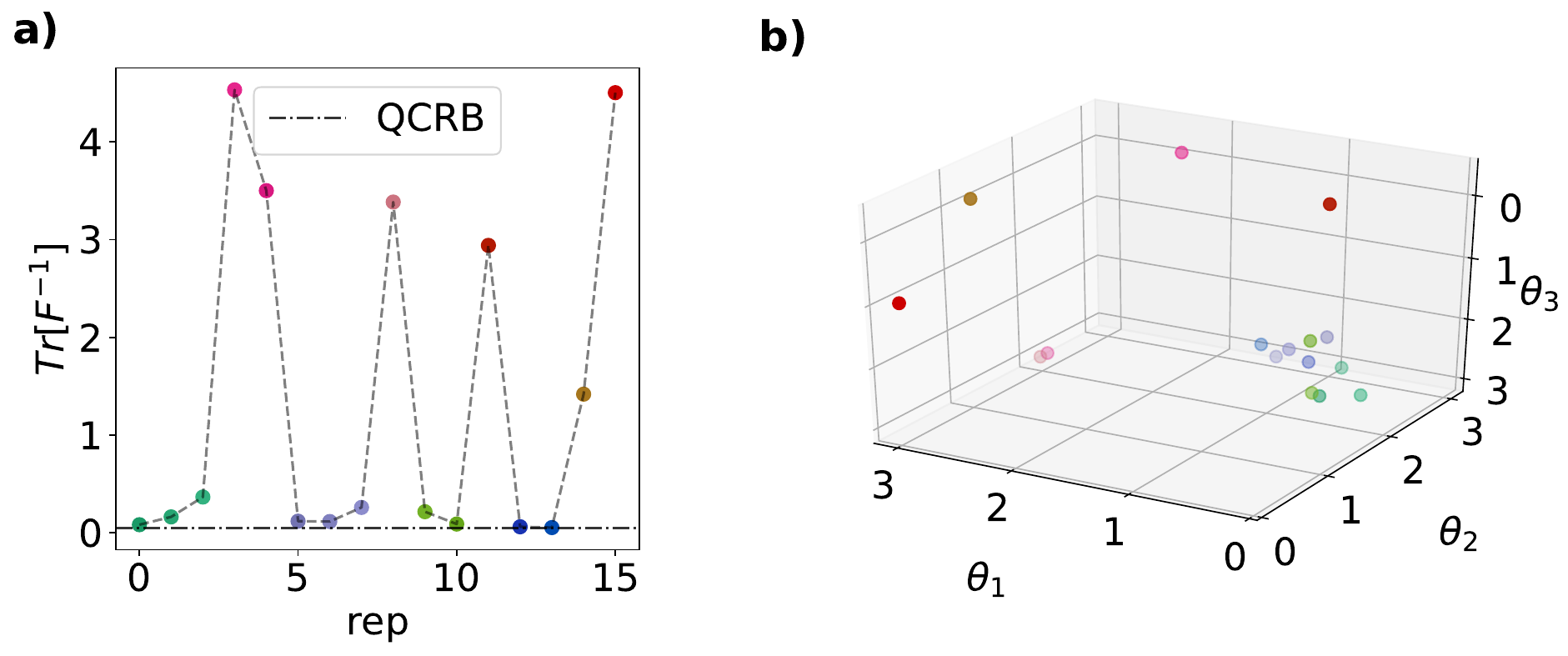}
\caption{Different repetitions (rep) of the variational optimization strategy. \textbf{a)} Value of the CRB reached for each optimization. \textbf{b)} Final configuration of the measurement settings found by the variational optimization. The color code links the $\{\theta_i\}$ settings with the corresponding minimum value of Tr$[F^{-1}]$ in panel a).
}
\label{fig:rep}
\end{figure}

Therefore, the choice of repeating the minimization strategy 3 times in order to select the proper experimental measurement settings $\{\theta_i\}$, allows also to drastically reducing the influence of such experimental fluctuations on the value to minimize. 
Indeed, repeating the variational optimization strategy multiple times it is evident that not always the global minimum of the function is reached. The results of this study for one configuration of the two-photon probes scenarios are reported in Fig.\ref{fig:rep}.

The measurement settings yielding the lowest cost function value, among the three repetitions, were selected as the optimal $\theta$-settings. This approach allowed us to increase the likelihood of converging to the global minimum of the cost function, thereby enhancing the reliability of the estimation process.

The estimation performances retrieved shifting the measurement settings in $\vec{\theta}_{opt}$ for $100$ repetitions of the estimate are reported in Fig.\ref{fig:FIrep} together with a Gaussian fit on the results.

\begin{figure}[ht!]
\centering
\includegraphics[width=0.6\textwidth]{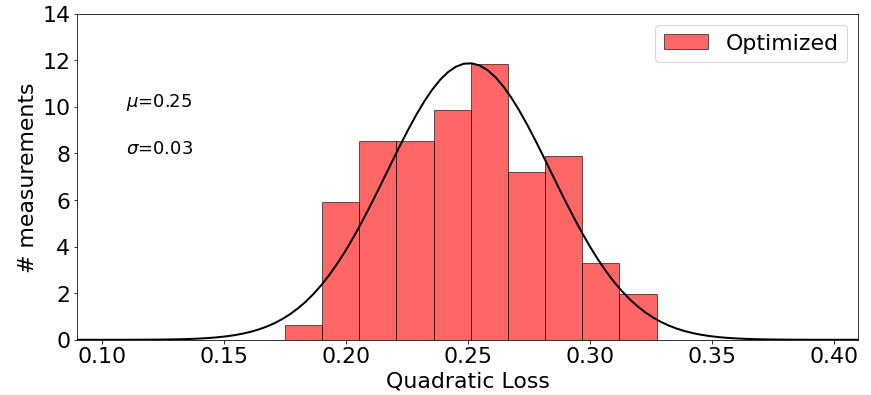}
\caption{Distributions of the experimental performances achieved on the configuration s10 when repeating the estimation procedure 100 times. The black line is a Gaussian fit on the results with mean $\mu$ and standard deviation $\sigma$.
}
\label{fig:FIrep}
\end{figure} 

\begin{figure}[ht!]
\centering
\includegraphics[width=0.6\textwidth]{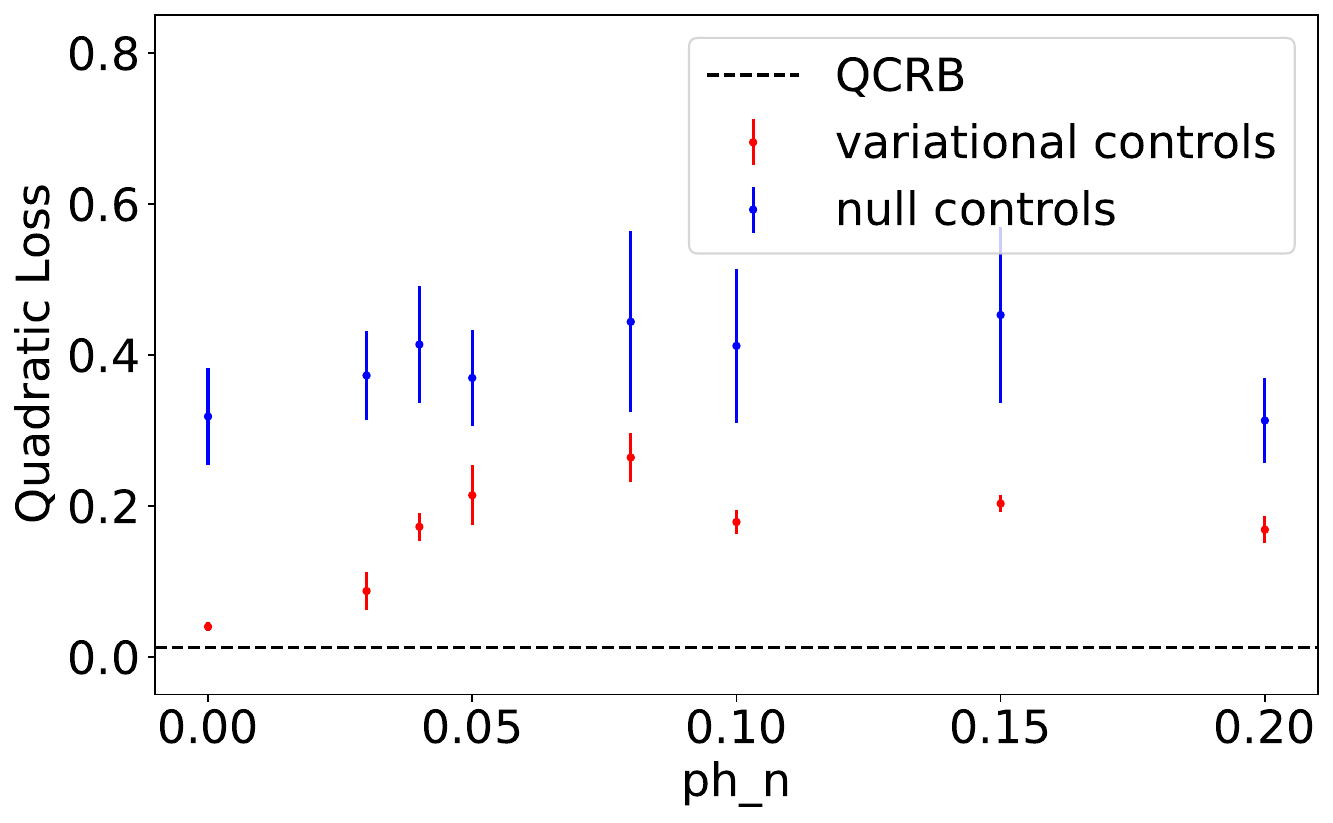}
\caption{Experimental quadratic loss obtained using single-photon probes for the estimate of the configuration $\overline{s}1$ with an additional level of phase noise $ph_n$. The results are the averages over 30 different repetitions of the estimate obtained without shifting the measurement settings (blu points) and setting the variational controls (red points).
}
\label{fig:ph_n}
\end{figure}

In conclusion, we report here the performances in terms of quadratic loss, of the experiment performed with an additional phase noise. In this scenario the true phases set does not correspond to the corresponding triplet $\vec{\phi}_{true}$ but shifting all the vector components by a quantity equal to $ph_n$. The average quadratic loss obtained after repeating the experiment 30 times seeding the system
with $M = 500$ single-photon probes are reported in Fig.\ref{fig:ph_n}. Also in such noisy conditions the strategy that makes use of the variational measurement settings shows enhanced performances compared to the non-shifted measurements scenario.

\providecommand{\noopsort}[1]{}\providecommand{\singleletter}[1]{#1}%